\documentclass{pasj00}
\usepackage[dvips]{color}

\begin{document}
\SetRunningHead{Y. Ita et al.}{AKARI observations of circumstellar dust in NGC104 and NGC362}
\Received{yyyy/mm/dd}
\Accepted{yyyy/mm/dd}

\title{AKARI observations of circumstellar dust in the globular clusters NGC104 and NGC362}

\author{Yoshifusa \textsc{Ita}%
}
\affil{$^1$Institute of Space and Astronautical Science, Japan Aerospace Exploration Agency \\ 3-1-1 Yoshinodai, Sagamihara, Kanagawa 229-8510, Japan}
\email{yita@ir.isas.jaxa.jp}

\author{Toshihiko \textsc{Tanab\'{e}}$^2$, Noriyuki \textsc{Matsunaga}$^2$, Yoshikazu \textsc{Nakada}$^2$, Mikako \textsc{Matsuura}$^3$,}
\affil{$^2$Institute of Astronomy, Graduate School of Science, \\ The University of Tokyo, 2-21-1 Osawa, Mitaka, Tokyo 181-0015, Japan}
\affil{$^3$National Astronomical Observatory of Japan, 2-21-1 Osawa, Mitaka, Tokyo, 181-8588, Japan}
\and
\author{Takashi \textsc{Onaka}$^4$, Hideo \textsc{Matsuhara}$^1$, Takehiko \textsc{Wada}$^1$, Naofumi \textsc{Fujishiro}$^5$, Daisuke \textsc{Ishihara}$^4$,}
\author{Hirokazu \textsc{Kataza}$^1$, Woojung \textsc{Kim}$^1$, Toshio \textsc{Matsumoto}$^1$, Hiroshi \textsc{Murakami}$^1$, Youichi \textsc{Ohyama}$^1$,}
\author{Fumihiko \textsc{Usui}$^1$, Shinki \textsc{Oyabu}$^1$ Itsuki \textsc{Sakon}$^4$, Toshinobu \textsc{Takagi}$^1$, Kazunori \textsc{Uemizu}$^1$,}
\author{Munetaka \textsc{Ueno}$^6$, Hidenori \textsc{Watarai}$^7$}
\affil{$^4$Department of Astronomy, Graduate School of Science, \\ The University of Tokyo, Bunkyo-ku, Tokyo 113-0033, Japan}
\affil{$^5$Department of Physics, Graduate School of Science, \\ The University of Tokyo, Bunkyo-ku, Tokyo 113-0033, Japan}
\affil{$^6$Department of Earth Science and Astronomy, Graduate School of Arts and Sciences,
 \\ The University of Tokyo, Meguro-ku, Tokyo 153-8902, Japan}
\affil{$^7$Office of Space Applications, Japan Aerospace Exploration Agency, \\ Tsukuba, Ibaraki 305-8505, Japan}


%


\maketitle

\begin{abstract}
We report preliminary results of AKARI observations of two globular clusters, NGC104 and NGC362. Imaging data covering areas of about 10 $\times$ 10 arcmin$^2$ centered on the two clusters have been obtained with InfraRed Camera (IRC) at 2.4, 3.2, 4.1, 7.0, 9.0, 11.0, 15.0, 18.0 and 24.0 $\mu$m. We used F$_{11}$/F$_{2}$ and F$_{24}$/F$_{7}$ flux ratios as diagnostics of circumstellar dust emission. Dust emissions are mainly detected from variable stars obviously on the asymptotic giant branch, but some variable stars that reside below the tip of the first-ascending giant branch also show dust emissions. We found eight red sources with F$_{24}$/F$_{7}$ ratio greater than unity in NGC362. Six out of the eight have no 2MASS counterparts. 
However, we found no such source in NGC104.
\end{abstract}

\section{Introduction}

In order to calibrate the geometric distortion of the InfraRed Camera (IRC; \cite{onaka}) onboard AKARI \citep{murakami}, we observed two galactic globular clusters, namely NGC104 ($=$ 47 Tuc) and NGC362 during the AKARI in-orbit performance verification phase. 
Four pointings were dedicated for these observations. 
Here, we will use these data to study the evolution of low-mass stars with an emphasis on mass loss from them.

Galactic globular clusters are the best test grounds for stellar evolution theories of low-mass stars, since we can reasonably assume that they are composed of a single stellar population -- that is, their constituent stars were formed at the same time, in the same volume of space, and from the same cloud of gas. Understanding of the evolution of low-mass stars is important because they may lose as much as $\sim$ 40\% of their initial mass during their life time (\cite{wachter2002}), and also, as they represent the majority of the stars in the Galaxy, their role in the galactic formation and evolution is not negligible (e.g., \cite{schroder2001}).

The mass loss process, which dominates the evolution of the star itself, is still a poorly understood phenomenon. Many observations (e.g., \cite{tanabe1997}) showed that the phase of the highest mass loss rate is achieved during the last stage of the asymptotic giant branch (AGB) evolutionary phase. Although it is not a large scale, however, mass loss also occurs along the first-ascending giant branch (RGB). It should exert a greater effect on stellar evolution, especially for low mass stars, given that their residence time in the RGB is much longer than that in the AGB (\cite{schroder2005}).

NGC104 and NGC362 are well suited for studying a mass loss history along the RGB and AGB, because of well populated RGB and AGB stars. AKARI/IRC can detect red giants well below the tip of the RGB in the two clusters. The basic parameters of NGC104 and NGC362 are listed in Table~\ref{table:parameter}, and a short description of each cluster including introductions of previous work follows.

\begin{table}[htbp]
  \caption{The basic parameters of NGC104 and NGC362.}\label{table:parameter}
  \begin{center}
    \begin{tabular}{lrr}
    \hline
                       & NGC104               & NGC362         \\
    \hline
    DM [mag]$^*$       & 13.50 $\pm$ 0.08$^1$ & 15.06 $^3$     \\
    $[$Fe/H$]$         & $-$0.66$^2$          & $-$1.12$^3$    \\
    Age [Gyr]          & 11.2 $\pm$ 1.1$^1$   & 8 $\sim$ 9$^3$ \\
    \hline
    \end{tabular}
  \end{center}
$^*$ Distance modulus. \\
References: $^1$ \citet{gratton2003}, $^2$ \citet{carretta1997}, $^3$ \citet{gratton1997}
\end{table}

\subsection{NGC104}
NGC104 is the second brightest (after $\omega$ Cen) galactic globular cluster. 
To date, 42 long period variables have been found (\cite{lebzelter2005}). Mid-IR surveys of this cluster have been done by using the ISOCAM \citep{cesarsky1996} on board the Infrared Space Observatory (ISO; \cite{kessler1996}) to study the evolution of dust mass loss along the RGB and AGB (e.g., \cite{ramdani2001}, \cite{origlia2002}). \citet{ramdani2001} observed outer regions of NGC104, and \citet{origlia2002} observed an area of about \timeform{5'} $\times$ \timeform{3'} centered on the cluster core. The IRC's wide field of view (\timeform{10'} $\times$ \timeform{10'}) enables us to cover both the central and outer regions within a single observation.

\subsection{NGC362}
This cluster is also well studied. Along with NGC288 it forms one of the most famous ``second parameter" couples. NGC362 and NGC288 have about the same metallicities ($=$ ``the first or main parameter"), but their horizontal-branch (HB) morphologies are different. NGC362 has a red HB morphology, while NGC288 has a blue one (e.g., \cite{sandage1960}, \cite{bolte1989}, \cite{green1990}). To explain the difference, there must be at least one "second parameter". There are many second parameter candidates, e.g., cluster age, mass loss along the RGB, helium abundance, rotation and deep helium mixing, dynamical interactions involving binaries and even planets, environmental effects in high-density environments, and so on (e.g., \cite{vink2002} and references therein), and its origin is still a controversy (e.g., \cite{stetson1996}). Mass loss episodes along the RGB may be an "inclusive" second parameter (\cite{catelan2001}), as they are derivative (i.e., determined by other parameters of the star). \citet{origlia2002} also observed this cluster, finding three stars with mid-infrared excess.

\section{Observations \& Reductions}
Imaging observations of NGC104 and NGC362 were obtained on 2006 May 1st (NGC104), and 6th and 7th (NGC362) UT with the IRC. The IRC03 AKARI IRC observing template (AOT03) was used, yielding imaging data at 2.4, 3.2, 4.1, 7.0, 9.0, 11.0, 15.0, 18.0 and 24.0 $\mu$m taken in at least 2 dithered positions.

Raw data were processed with the IRC imaging data pipeline, version 070104 (see IRC Data User's Manual \citet{lorente2007} for details). The resultant IRC mosaic images have pixel sizes of \timeform{1.46''} pixel$^{-1}$, \timeform{2.40''} pixel$^{-1}$ and \timeform{2.38''} pixel$^{-1}$ for NIR (2.4, 3.2, and 4.1 $\mu$m), MIR-S (7.0, 9.0, and 11.0 $\mu$m) and MIR-L (15.0, 18.0 and 24.0 $\mu$m) channels of IRC, respectively, covering an area $\sim$ 100 arcmin$^2$ around the cores of NGC104 and NGC362 in each wavelength.

\subsection{Photometry}
To derive calibrated fluxes for each star, point spread function (PSF) fitting photometry was performed on the mosaiced images with the IRAF\footnote{IRAF is distributed by the National Optical Astronomy Observatories, which are operated by the Association of Universities for Research in Astronomy, Inc., under cooperative agreement with the National Science Foundation.} package DAOPHOT. Photometry was done for each mosaiced image independently. This involved the following steps:
\begin{enumerate}
  \item DAOFIND was used to find stars whose fluxes are at least 5 $\sigma$ above the background, where $\sigma$ is the background noise estimated locally around stars.
  \item Aperture photometry was performed on all of the stars found in step 1, using the task PHOT with aperture radii of 10.0 and 7.5 pixels for NIR and MIR$-$S/MIR$-$L images, respectively. We used the same aperture radii as had been used in the standard star flux calibration (Tanab\'{e} et al. in preparation), so the aperture corrections were not applied. The resultant astronomical data units were converted to the calibrated fluxes by using the IRC flux calibration constants version 070119. 
  \item Several stars with moderate flux (i.e., with a good signal-to-noise ratio and unsaturated) and without neighbors within 7 pixels were selected from the results of step 2. We found more than 5 such stars in each mosaiced image. The selected stars were used to construct a model PSF.
  \item The PSF fit was adopted to all of the stars found in the mosaiced images using ALLSTAR to get their instrumental fluxes and their corresponding errors. To check the array-location-dependence, we ran ALLSTAR with an option that the PSF can be linearly variable over the images. With this test, we found that the PSF does not vary significantly over the array. Hence constant PSF is assumed over an image.
  \item The resultant instrumental fluxes were shifted so that the instrumental fluxes of the stars selected in step 3 match the calibrated fluxes calculated in step 2. 
\end{enumerate}

We do not deredden the measured fluxes, but we applied color corrections on the calibrated fluxes by assuming a black body with the effective temperature of 3500 K. The temperature change of $\pm$ 500 K yields 2.4\% change in the correction factor in the 2.4 $\mu$m case, but less than 1.0\% for the others. Therefore, any discussions followed are almost insensitive to the assumed reference black body temperature.

\subsection{Cross-identification with 2MASS sources}
\begin{figure}[ht]
  \begin{center}
    \FigureFile(78mm,78mm){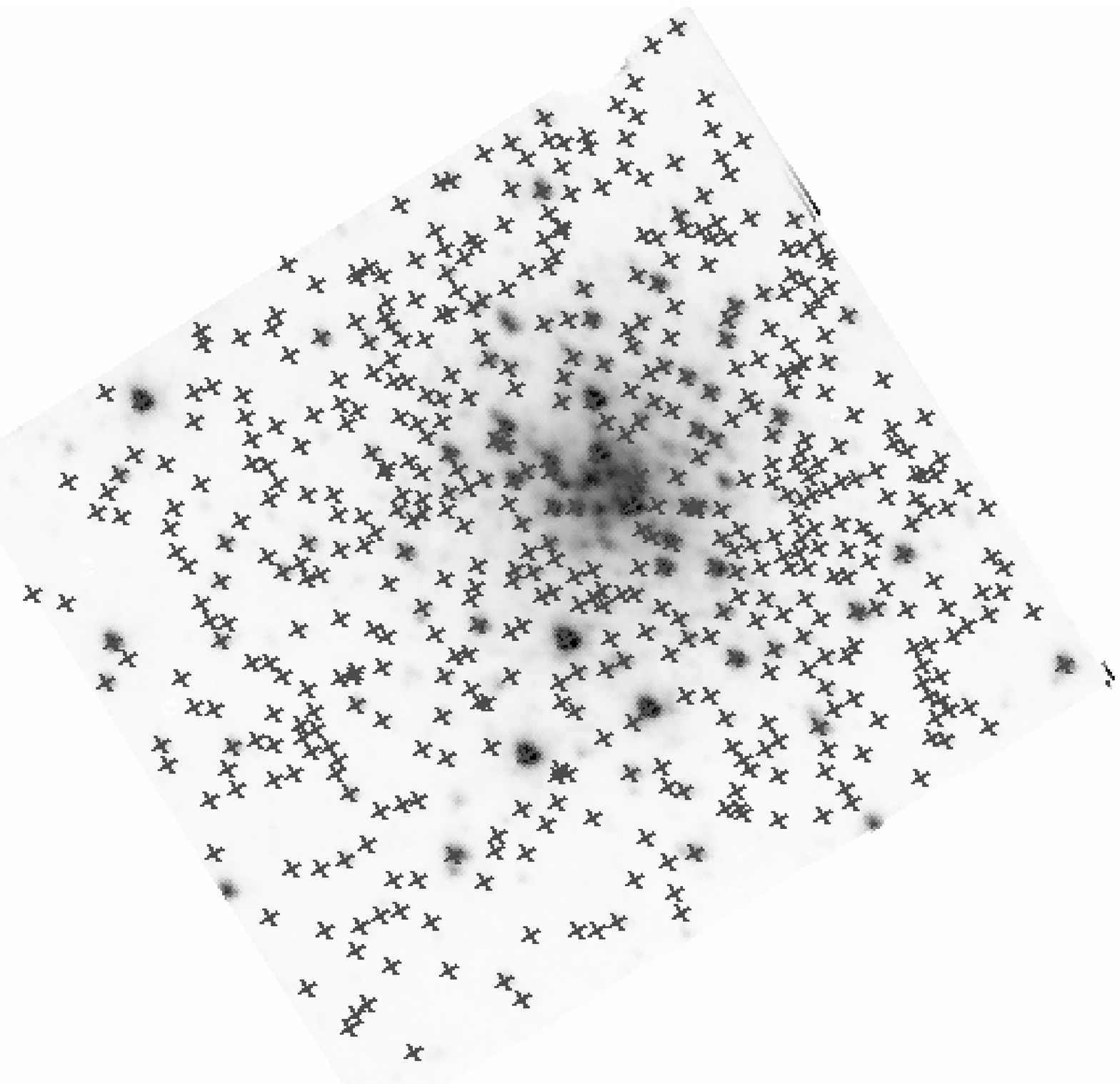}
    \FigureFile(78mm,78mm){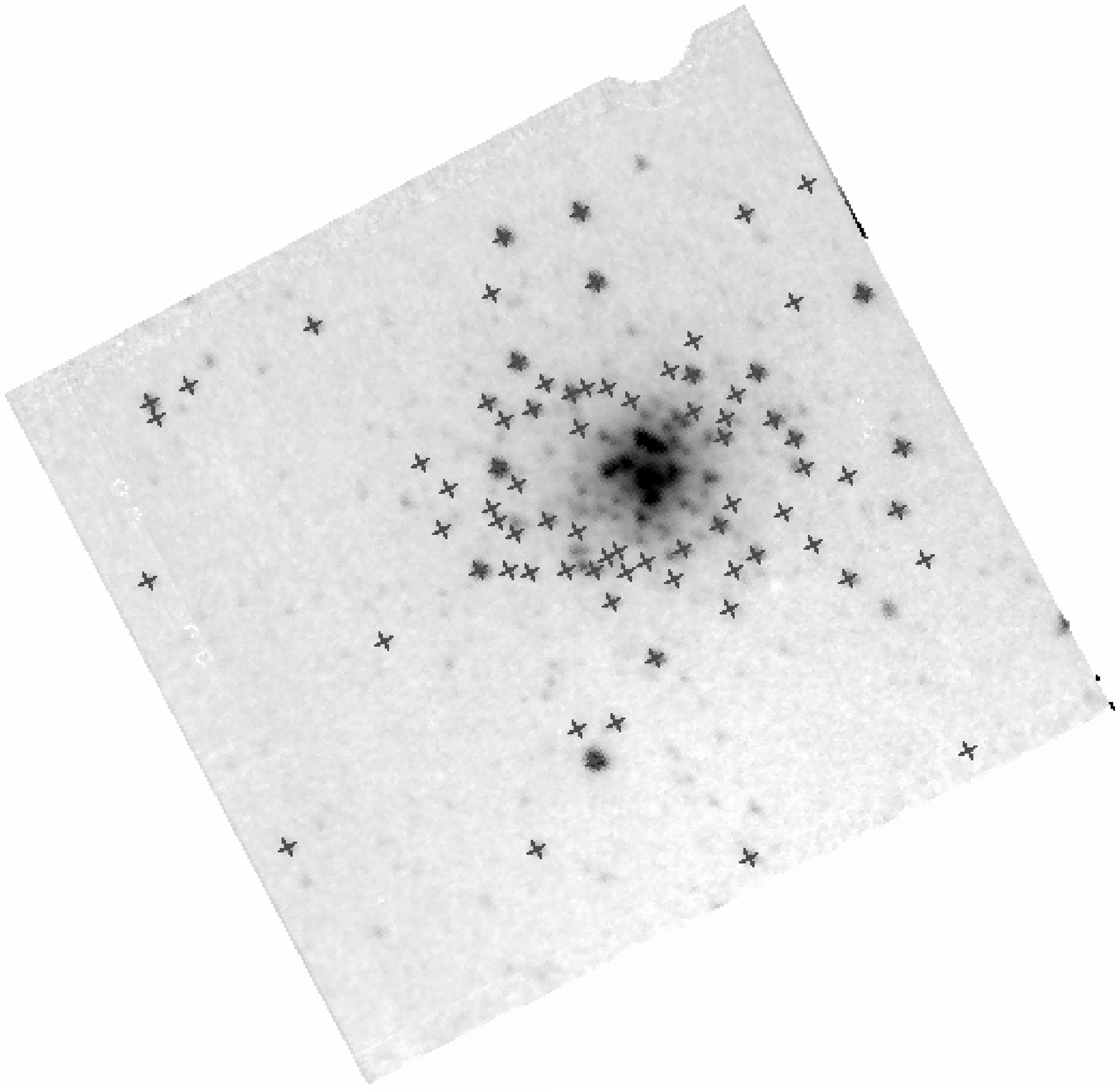}
  \end{center}
  \caption{AKARI IRC 11.0 $\mu$m image of NGC104 (top) and NGC362 (bottom). Crosses show IRC 11.0 $\mu$m sources with 2MASS counterpart. North is up, and east is to the left.}
  \label{fig:ngc104-ngc362}
\end{figure}
Cross-correlation of the IRC sources with 2MASS sources has been made in order to determine the astrometric coordinates of detected sources in the mosaiced images of each wavelength. As the result, their coordinates were determined to the accuracy of about \timeform{2.0''} relative to the corresponding 2MASS sources. In Figure \ref{fig:ngc104-ngc362}, we show the results of the cross-correlation between AKARI IRC 11.0 $\mu$m and 2MASS sources. The background images are IRC 11.0 $\mu$m images of NGC104 (top) and NGC362 (bottom). Crosses represent the IRC 11.0 $\mu$m sources with 2MASS counterparts within a radius of \timeform{1.0''}.

After this cross-correlation, we inspected each source with $K_{\textrm{2MASS}}$ $<$ 13.0 mag by the eye to remove miss-identifications. Since we worked with the photometric data in high stellar density regions, we eliminated any IRC sources that were merged (we assumed that any IRC sources that have neighbor(s) within a half width at half maximum of the PSF of each wavelength as merged) with a brighter star, because they would introduce elements of confusion into the following discussions.

\begin{figure}[htbp]
  \begin{center}
    \FigureFile(81.5mm,81.5mm){figure2.ps}
  \end{center}
  \caption{The F$_{11}$/F$_{2}$ vs $K_s$ diagram for sample stars with the variable stars identified. We use the nomenclature of \citet{lebzelter2005} for NGC104, and \citet{szekely2007} for NGC362, except for two stars C2 and C16 (see text). The F$_{2}$ flux densities were calculated from $K_s$ magnitude by adopting the zero-magnitude flux of 666.7 Jy (\cite{cohen2003}). See section 3.3 for the two triangles in NGC362. Closeup around the crowded part is shown in the inset.}
  \label{fig:f11vsf2}
\end{figure}

\section{Results \& Discussion}
\subsection{Infrared excess of cluster stars}
\citet{ramdani2001} showed that the ratio of the ISO 11.5 $\mu$m to DENIS $K_s$ 2 $\mu$m flux density is a good indicator of dust mass loss. Therefore, we made a IRC 11.0 $\mu$m to 2MASS $K_s$ flux density ratio (F$_{11}$/F$_{2}$) vs $K_{\textrm{2MASS}}$ diagram of our sample to see which stars show circumstellar dust emission. Hereafter, we denote 2MASS $K_s$ as $K_s$ unless otherwise described. The top panel of Figure \ref{fig:f11vsf2} is plotted for NGC104, and the bottom for NGC362. The $K_s$ fluxes (F$_{2}$) were calculated from $K_s$ magnitudes by adopting the zero-magnitude flux of 666.7 Jy (\cite{cohen2003}). We did not apply reddening correction on 2MASS nor IRC fluxes. To get a basic idea, we calculated the F$_{11}$/F$_{2}$ ratio expected for a dust-free stellar atmosphere by using ATLAS9 \citep{kurucz1993, sbordone2004, sbordone2005}. It came out that a star of T$_{\textrm{eff}} = 3500$ K, log $g$ $= 1.5$, v$_{\textrm{turb}}$ $= 2.0$ km/s, and $[\textrm{Fe/H}]$ $= -1.0$ would have F$_{11}$/F$_{2}$ ratio of about 0.065, showing that the F$_{11}$/F$_{2}$ ratios for most of the stars in NGC104 and NGC362 are consistent with the value predicted for dust-free photospheres. Further experiments with ATLAS9 showed that metallicity has a negligible impact on the F$_{11}$/F$_{2}$ ratio.

As it has been suggested by several authors (e.g., \cite{vassiliadis1993}), stellar pulsations  play a key role to trigger dust mass loss. The present results also show that all of the bright red giants with F$_{11}$ excess (F$_{11}$/F$_{2}$ $\ge 0.1$) are indeed variables. Throughout this paper, we use the nomenclature of \citet{lebzelter2005} for variable stars in NGC104, and of \citet{szekely2007} for ones in NGC362, except for C2 and C16. We have to note that the variable stars C2 and C16 in NGC362 are not listed in \citet{szekely2007}, but they are identical to V2 and V16 listed in \citet{clement2001}. It is confirmed by a near-infrared monitoring survey conducted by one of us (N. Matsunaga) that C2 is a semi-regular variable with a pulsation period of about 90 days, and also that C16 is a Mira-like variable with a pulsation period of about 135 days (\cite{sawyer1931, matsunagaD}).

Two sources in NGC104 with $K_{s} \sim 12.0$ and one source in NGC362 with $K_{s} \sim 13.4$ may have F$_{11}$/F$_{2}$ ratios greater than 0.1, but their photometric errors prevent us from regarding them as sources with infrared excess. Flux determinations of these three stars need further examination because they have much larger error bars than other stars at similar magnitudes. Put it all together, we found eleven stars (V1, V2, V3, V4, V8, V13, V21, LW7, LW10, LW11, and LW13) and two stars (C2 and C16) with infrared excess (F$_{11}$/F$_{2}$ $\ge$ 0.1) in our observed fields of NGC104 and NGC362, respectively.

\subsection{Infrared excess from RGB stars in NGC104?}
Figure \ref{fig:f11vsf2} also illustrates that a few variable stars below the tip of the first red giant branch (TRGB) in NGC104 do exhibit infrared excess. The TRGB occurs at $K=$ 6.75 $\sim$ 7.1 mag in NGC104 (\cite{ferraro2000, lebzelter2005}), and specifically, V13, LW7, and LW11 show large (F$_{11}$/F$_{2}$ $\ge 0.1$) infrared excess although they reside below the TRGB. These three stars have similar pulsation periods of about 40 days and also similar ($J-K$) colors of about 1.0 (\cite{lebzelter2005}). The period and ($J-K$) color are both being typical of variable stars below the TRGB found in the Large and Small Magellanic Clouds \citep{kiss2003, ita2004a, ita2004b}. As in \citet{ita2002}, stars below the TRGB could be either on the AGB or the RGB, but a substantial fraction could be RGB stars. The definite identification of the evolutionary stage of these stars is difficult based solely on the present data. We cannot rule out the possibility that they are thermally-pulsating AGB stars during the phase of quiescent helium burning, when stars are fainter by just over a magnitude (\cite{marigo2003}).
Further observations of each star is definitely needed for detailed study.

\begin{figure}[htbp]
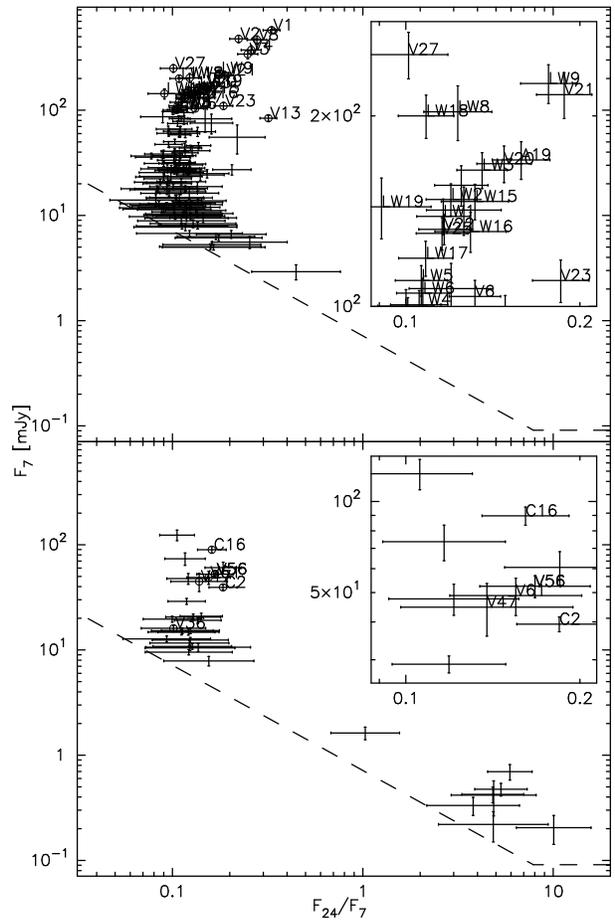

  \begin{center}
    \FigureFile(80mm,80mm){figure3.ps}
  \end{center}
  \caption{The F$_{24}$/F$_{7}$ vs F$_{7}$ diagram for sample stars with the variable stars identified as in figure \ref{fig:f11vsf2}. The dashed lines shows 5 $\sigma$ sensitivity limit for AOT03 in one pointed observation, as given in \cite{onaka}. Note that fluxes are color-corrected, but not dereddened. Closeup around the crowded part is shown in the inset.}
  \label{fig:f7vsf24}
\end{figure}

\begin{figure}[htbp]
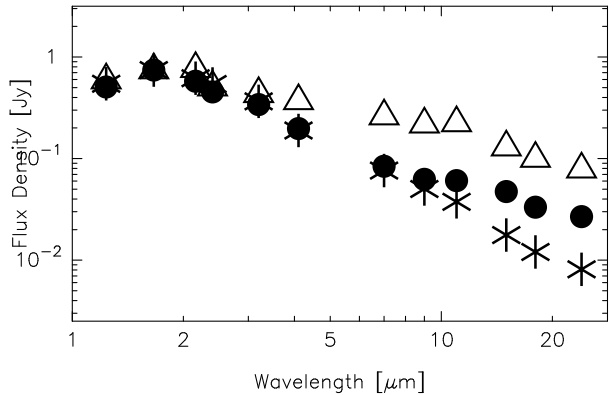

  \begin{center}
    \FigureFile(80mm,80mm){figure4.ps}
  \end{center}
 \caption{The spectral energy distribution of V13 (filled circles), 2MASS 00234761$-$7202498 (asterisks), and V1 (triangles) in NGC104. Note that the fluxes of 2MASS 00234761$-$7202498 and V1 are scaled so that the $H-$band flux densities of them are matched to that of V13. Photometric errors are smaller than the size of the marks.}
 \label{fig:v13}
\end{figure}

Interestingly, V13 in NGC104 has a F$_{24}$/F$_{7}$ ratio comparable to that of V1 (see the top panel of Figure \ref{fig:f7vsf24}), which has the longest pulsation period, largest pulsation amplitude, and highest luminosity among all the members of NGC104, and shows the typical silicate dust feature in its mid-IR spectrum with a mass-loss rate of $\sim$ 10$^{-6}$ \MO yr$^{-1}$ (\cite{vanloon2006}). We show the spectral energy distribution (SED) of V13 in Figure \ref{fig:v13} using 2MASS $JHK_s$ and all of the IRC fluxes. To emphasize the uniqueness of V13 compared to other normal red giants and also to genuine mass-losing AGB stars, the SED of a red giant, namely 2MASS 00234761-7202498 in NGC104 that has similar $K_s$ magnitude and $(J-K_s)$ color ($K_s=$ 7.606 mag, and $J-K_s$ = 1.081 mag) as those of V13 ($K_s=$ 7.755 mag, and $J-K_s$ = 1.089 mag), and also V1 are included in the figure. For comparison, the fluxes of 2MASS 00234761-7202498 and V1 are multiplied by a factor of 1.022 and 0.410, respectively, that were calculated as the $H-$band flux densities of V13, 2MASS 00234761-7202498, and V1 are matched. We also calculated the absolute bolometric magnitude of V13 by fitting two black body curves with T$_{\textrm{eff}} = 3261$ K and T$_{\textrm{eff}} = 263$ K to the observed data. We obtained $M_{\textrm{bol}} \sim -$3.11 mag as a result, using the distance modulus of 13.5 mag for NGC104 (see Table \ref{table:parameter}). \citet{lebzelter2006} took low-resolution mid-infrared (7.6$-$21.7 $\mu$m) spectra of V13 with the Spitzer telescope. They showed that V13 is devoid of a 9.7 $\mu$m emission band feature of amorphous silicate, but it has broad emission features at 11.5 $\mu$m (likely to be Al$_2$O$_3$), 13 $\mu$m (likely to be an Al$-$O stretching vibration), and 20 $\mu$m (no firm identification). Our results are consistent with theirs, showing that the infrared excess is detected at longward of 11.0 $\mu$m (Figure \ref{fig:v13}). Aluminium oxide features have been detected from low mass loss rate oxygen-rich AGB stars (\cite{onaka1989, kozasa1997}). Combined with the fact that the F$_{11}$/F$_{2}$ ratio of V13 is not so large, these results show that dust composition of V13 is different from those of usual mass losing AGB stars. 

\subsection{Very red sources in NGC362}
\begin{figure}[htbp]
  \begin{center}
    \FigureFile(80mm,80mm){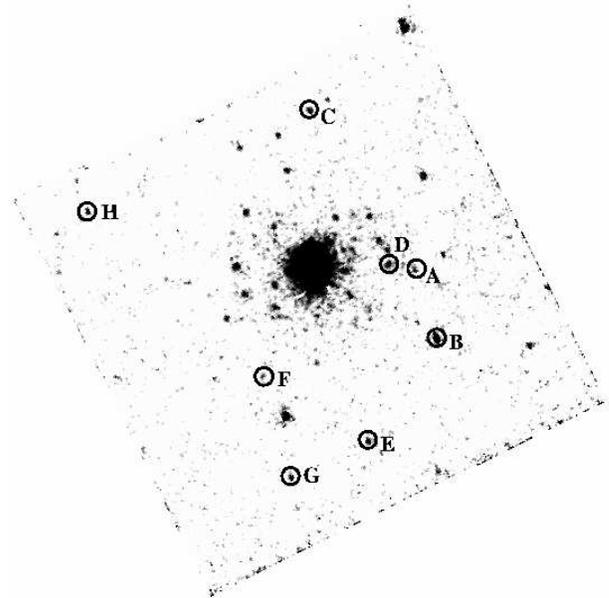}
  \end{center}
  \caption{AKARI IRC 24.0 $\mu$m image of NGC362 with eight red sources (F$_{24}$/F$_{7}$ $> 1.0$, see text) identified. their spectral energy distributions are shown in Figure \ref{fig:redsed}. North is up, and east is to the left.}
  \label{fig:spatial}
\end{figure}

\citet{boyer2006} observed M15 with IRAC \citep{fazio} and MIPS (\cite{rieke}) onboard Spitzer Space Telescope (\cite{werner}). Their F$_{24}$/F$_{8}$ vs F$_{8}$ diagram revealed that there are at least 23 red sources in M15. They suggested that the red sources are mass-losing AGB or post-AGB candidates after consideration of their loose spatial distribution,

We use our 7.0 and 24.0 $\mu$m data to make the F$_{24}$/F$_{7}$ vs F$_{7}$ diagram, as shown in Figure \ref{fig:f7vsf24}. The top panel is for NGC104 and the bottom is for NGC362. It is seen that there are no sources with F$_{24}$/F$_{7}$ $\ge 1.0$ in NGC104, but there are eight red sources in NGC362. Six out of the eight have no 2MASS counterparts within a radius of \timeform{6''}. The other two sources with 2MASS counterparts have large F$_{11}$/F$_{2}$ ratios, as seen in Figure \ref{fig:f11vsf2} (triangles). The spatial distribution of the eight sources are shown in Figure \ref{fig:spatial}, indicating that they are distributed around the cluster, and are not biased to the cluster center. 

\begin{figure}[htbp]
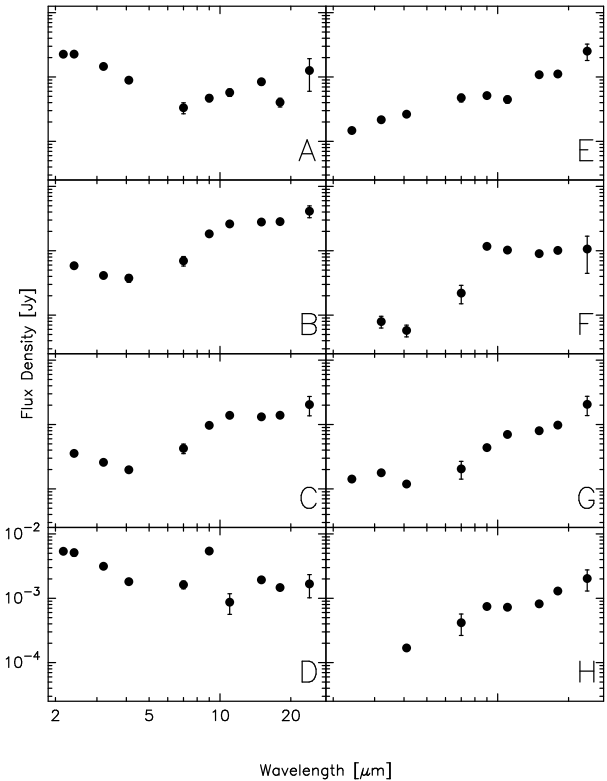

  \begin{center}
    \FigureFile(80mm,80mm){figure6.ps}
  \end{center}
  \caption{Spectral energy distributions of the eight red sources in NGC362 showing $K_s$ (if available) and all of the AKARI/IRC data. Note that some stars were not detected even in the NIR channel of IRC. The scales on the x- and y-axis are the same for each panel. The labels are as in Figure \ref{fig:spatial}.}
  \label{fig:redsed}
\end{figure}

According to a model calculation (Pearson et al. in preparation), we can expect one or two galaxies in the IRC 100 arcmin$^2$ field of view down to the 5 $\sigma$ sensitivity limit of 24 $\mu$m. Then, there is little possibility that these sources are all background galaxies. Absence of red sources in NGC104 also suggests it unlikely that they are all galaxies. It is also unlikely that they are mass-losing AGB stars in NGC362, because their F$_{7}$ flux densities are too faint \citep{groenewegen2006}. \citet{boyer2006} found similar objects in M15 by Spitzer observations and pointed out the possibility that they could be post-AGB stars. However, the fact that no clear counterparts have been seen at NIR wavelegnths may be incompatible with the post-AGB identification. NGC362 lies near the Small Magellanic Cloud in projection, therefore they might be bright high-mass-losing AGB stars in the SMC. However, such stars should be very rare, and we would not expect to detect eight of such stars in a 100 arcmin$^2$ field. We show the spectral energy distributions of the eight sources in Figure \ref{fig:redsed}. It can be seen that infrared excess is detected not only at 24.0 $\mu$m but also at 15.0 and 18.0 $\mu$m. Therefore it is likely that the excess is continuum emission and that the contribution from line emission such as [O IV] 26 $\mu$m is insignificant. Some spectra show peculiar features, but the presence of the excess seems to be secure (based on the multi-band photometry). To identify the eight red sources and also to confirm the photometric results, infrared spectroscopic observations with AKARI and/or Spitzer would be needed.

\section{Summary}
We presented the first AKARI/IRC imaging data of NGC104 and NGC362. We found that stars with large F$_{11}$/F$_{2}$ ratio are all variables, confirming the close link between mass loss and stellar pulsation. We detected eight sources with large F$_{24}$/F$_{7}$ ratio in NGC362, but no such sources were present in NGC104. 
We also showed that V13 in NGC104 has interesting features, as being a possible RGB candidate with infrared excess.

\section*{Acknowledgements}
We thank the referee for his/her useful comments which helped us improve this paper. We would like to thank Chris Pearson for making his latest galaxy count model calculations available in advance of publication. We also thank Michael Feast for helpful comments on the first version of the manuscript. AKARI is a JAXA project with the participation of ESA. This work is supported by the Grant-in-Aid for Encouragement of Young Scientists (B) No.~17740120 from the Ministry of Education, Culture, Sports, Science and Technology of Japan. This publication makes use of data products from the Two Micron All Sky Survey, which is a joint project of the University of Massachusetts and the Infrared Processing and Analysis Center, Caltech, funded by the National Aeronautics and Space Administration and the National Science Foundation.



\begin{thebibliography}{}
\bibitem[Bolte (1989)]{bolte1989}
   Bolte M. \ 1989, AJ, 97, 1688
\bibitem[Boyer et al. (2006)]{boyer2006}
   Boyer M.L, Woodward C.E., van Loon J.Th, et al. \ 2006, AJ, 132, 1415
\bibitem[Carretta and Gratton (1997)]{carretta1997}
   Carretta E., Gratton R.G. \ 1997, A\&AS, 121, 95
\bibitem[Catelan et al. (2001)]{catelan2001}
   Catelan M., Bellazzini M., Landsman W.B., \ 2001, AJ, 122, 3171
\bibitem[Cesarsky et al. (1996)]{cesarsky1996}
   Cesarsky C., Abergel A., Agnese P., et al. \ 1996, A\&A, 315, L32
\bibitem[Clement et al. (2001)]{clement2001}
   Clement C.M., Muzzin A., Dufton Q. \ 2001, AJ, 122, 2587 
\bibitem[Cohen et al. (2003)]{cohen2003}
   Cohen M., Wheaton W. A., \& Megeath S.T. \ 2003, AJ, 126, 1090
\bibitem[Fazio et al. (2004)]{fazio}
   Fazio G.G., Hora J.L., Allen L.E., et al. \ 2004, ApJS, 154, 10
\bibitem[Ferraro et al. (2000)]{ferraro2000}
   Ferraro F.R., Montegriffo P., Origlia L., et al. \ 2000, AJ, 119, 1282
\bibitem[Gratton et al. (1997)]{gratton1997}
   Gratton R.G., Fusi Pecci F., Carretta E., et al. \ 1997, astro-ph/9707107   
\bibitem[Gratton et al. (2003)]{gratton2003}
   Gratton R.G., Bragaglia A., Carretta E., et al. \ 2003, A\&A, 408, 529
\bibitem[Green and Norris (1990)]{green1990}
   Green E.M., Norris J.E. \ 1990, ApJ, 353, L117
\bibitem[Groenewegen (2006)]{groenewegen2006}
   Groenewegen M.A.T \ 2006, A\&A, 448, 181
\bibitem[Ita et al. (2002)]{ita2002}
   Ita Y., Tanab\`{e} T., Matsunaga N., et al. \ 2002, MNRAS, 337, L31
\bibitem[Ita et al. (2004a)]{ita2004a}
   Ita Y., Tanab\`{e} T., Matsunaga N., et al. \ 2004a, MNRAS, 347, 720
\bibitem[Ita et al. (2004b)]{ita2004b}
   Ita Y., Tanab\`{e} T., Matsunaga N., et al. \ 2004b, MNRAS, 353, 705
\bibitem[Kessler et al. (1996)]{kessler1996}
   Kessler M.F., Steinz J.A., Anderegg M.E., et al. \ 1996, A\&A, 315, L27
\bibitem[Kiss and Bedding (2003)]{kiss2003}
   Kiss L.L, Bedding T.R., \ 2003, MNRAS, 343, 79
\bibitem[Kozasa and Sogawa (1997)]{kozasa1997}
   Kozasa T., Sogawa H., \ 1997, Ap\&SS, 251, 165
\bibitem[Kurucz (1993)]{kurucz1993}
   Kurucz R.L. \ 1993, ATLAS9 Stellar Atmosphere Programs and 2 km/s grid. Kurucz CD-ROM No. 13. Cambridge, Mass.: Smithsonian AstrophysicalObservatory, 1993., 13
\bibitem[Lebzelter \& Wood (2005)]{lebzelter2005}
   Lebzelter T., Wood P.R. \ 2005, A\&A, 441, 1117
\bibitem[Lebzelter et al. (2006)]{lebzelter2006}
   Lebzelter T., Posch T., Hinkle K., et al., \ 2006, ApJ, 653, L145
\bibitem[Lorente et al. (2007)]{lorente2007}
   Lorente R., Onaka T., Ita Y., et al., \ 2007, AKARI IRC Data User's Manual ver. 1.1, http://www.ir.isas.jaxa.jp/AKARI/Observation/
\bibitem[Marigo et al. (2003)]{marigo2003}
   Marigo P., Girardi L., Chiosi C., \ 2003, A\&A, 403, 225
\bibitem[Matsunaga (2007)]{matsunagaD}
   Matsunaga N, \ 2007, Ph.D. thesis, University of Tokyo
\bibitem[Murakami et al. (2007)]{murakami}
   Murakami H., et al., \ 2007, PASJ, submitted
\bibitem[Onaka et al. (1989)]{onaka1989} 
   Onaka T., De Jong T., Willems F.J., \ 1989, A\&A, 218, 1690
\bibitem[Onaka et al. (2007)]{onaka} 
   Onaka T., et al., \ 2007, PASJ, submitted
\bibitem[Origlia et al. (2002)]{origlia2002} 
   Origlia L., Ferraro F.R., Pecci F.F., et al. \ 2002, ApJ, 571, 458
\bibitem[Ramdani and Jorissen (2001)]{ramdani2001}
   Ramdani A., Jorissen A. \ 2001, A\&AS 372, 85
\bibitem[Rieke et al. (2004)]{rieke}
   Rieke G.H., Young E.T., Engelbracht C.W., et al. \ 2004, ApJS, 154, 25
\bibitem[Sandage and Wallerstein (1960)]{sandage1960}
   Sandage A., Wallerstein G. \ 1960, ApJ, 131, 598
\bibitem[Sawyer (1931)]{sawyer1931}
   Sawyer, H.B, \ 1931, Harvard College Observatory Circular, vol. 366, pp.1-36 
\bibitem[Sbordone et al. (2004)]{sbordone2004}
   Sbordone L., Bonifacio P., Castelli F. \& Kurucz, R.L. \ 2004, Memorie della Societa Astronomica Italiana Supplement, 5, 93
\bibitem[Sbordone (2005)]{sbordone2005}
   Sbordone L. \ 2005, Memorie della Societa Astronomica Italiana Supplement, 8, 61
\bibitem[Schr\"{o}der and Sedlmayr (2001)]{schroder2001}
   Schr\"{o}der K.-P., Sedlmayr E. \ 2001, A\&A, 366, 913
\bibitem[Schr\"{o}der and Cuntz (2005)]{schroder2005}
   Schr\"{o}der K.-P., Cuntz M. \ 2005, ApJ, 630, L73
\bibitem[Stetson et al. (1996)]{stetson1996}
   Stetson P.B., VandenBerg D.A. \& Bolte M.J. \ 1996, PASP, 108, 560
\bibitem[Szek\'{e}ly et al. (2007)]{szekely2007}
   Szekely P., Kiss L.L., Jackson R., et al. \ 2007, astro-ph/0611663
\bibitem[Tanab\'{e} et al. (1997)]{tanabe1997}
   Tanabe T., Nishida S., Matsumoto S., et al. \ 1997, Natur, 385, 509
\bibitem[van Loon et al. (2006)]{vanloon2006}
   van Loon J.Th, McDonald M., Oliveira J.M., et al. \ 2006, A\&A, 450, 339
\bibitem[Vassiliadis and Wood (1993)]{vassiliadis1993}
   Vassiliadis E., Wood P.R. \ 1993, ApJ, 413, 641
\bibitem[Vink and Cassisi (2002)]{vink2002}
   Vink J.S., Cassisi S., \ 2002, A\&A, 392, 553
\bibitem[Wachter et al. (2002)]{wachter2002}
   Wachter A., Schr\"{o}der K.-P., Winters J.M., et al. \ 2002, A\&A, 384, 452
\bibitem[Werner et al. (2004)]{werner}
   Werner M.W., Roellig T.L., Low F.J., \ 2004, ApJS, 154, 1
\end{thebibliography}
\end{document}